# Inelastic neutron scattering: A novel approach towards determination of equilibrium isotopic fractionation factors. Size effects on heat capacity and beta-factor of diamond

Andrey A. Shiryaev[1,*], Veniamin B. Polyakov[2,**], Stephane Rols[3], Antonio Rivera[4],

Olga Shenderova[5]

1. A.N. Frumkin Institute of physical chemistry and electrochemistry RAS, Leninsky pr. 31 korp. 4, 119071, Moscow

2. D.S.Korzhinskii Institute of Experimental Mineralogy RAS, Academica Osypyana ul., 4, Chernogolovka, Moscow region, 142432, Russia

3. Institut Laue-Langevin, 71 avenue des Martyrs, CS 20156, 38042 Grenoble, Cedex 9, France

4. Instituto de Fusión Nuclear, Dep. Ingeniería Energética, ETSI Industriales, Universidad Politécnica de Madrid, Spain

5. Adámas Nanotechnologies, 8100 Brownleigh Drive, Raleigh, NC 2761, USA

* Corresponding author: Shiryaev A.A., email: a_shiryaev@mail.ru AND shiryaev@phyche.ac.ru

** Corresponding author: Polyakov V.B., email: polyakov@iem.ac.ru AND vpolyakov@mail.ru

## Abstract

A new experimental method of determination of equilibrium isotopic properties of substances based on Inelastic Neutron Scattering (INS) is proposed. We present mathematical formalism allowing calculation of beta-factor of single-element solids based on INS-derived Phonon Density of States (PDOS). PDOS data for nanodiamonds of widely different sizes and of macroscopic diamond were determined from Inelastic Neutron Scattering experiment. This allowed determination of heat capacities and, for the first time, β-factors for the diamond nanoparticles. We demonstrate considerable size-dependent increase of the heat capacities and decrease of the beta-factors for nanodiamonds relative to bulk diamond. Contributions of surface impurities/phases and phonon confinement to the size effects are evaluated. Applications to formation of diamond nanoparticles in nature are briefly discussed.

## 1. INTRODUCTION

Nanoparticles are ubiquitous in various natural environments. For example, carbonaceous and oxide nano- and micrograins are formed in large amounts in stellar outflows and comprise important fraction of interstellar medium[1]. However, thermodynamic data for nanocarbons and for nanoparticles in general are scarce, despite their importance for correct interpretation of observations and modelling. It is important to stress that not only mere physical dimensions of a phase give rise to the size or confinement effect. For example, high concentration of point and extended defects may dramatically change electronic and thermal vibration properties of matter giving rise to the confinement effects in relatively large particles and even in bulk materials.

Knowledge of heat capacity of nanoparticles is important for cosmochemistry and astrophysics. An isolated nanoparticle struck by an UV photon[2] or an ion[3,4] may experience very significant temperature excursion, so-called stochastic heating. This process may change structure of the nanoparticle (e.g., induce conversion of diamond phase to $sp^2$-carbon), anneal defects, etc. One can expect that the heating will be more significant for the smallest particles. Calculations of magnitude of the heating require, besides other things, knowledge of thermal capacity.

The insufficiency of thermodynamic data is even more glaring in case of isotope properties of nanoparticles. Distribution of stable isotopes between different phases and minerals contains important and often unique information about geochemical processes in the Earth and planetary interiors. Knowledge of equilibrium stable isotope fractionation factors is a key instrument for correct interpretation of geochemical information encoded in the observed stable isotope distributions. However, direct measurements of the equilibrium fractionation factors in stable isotope exchange experiments require attainment of the high degree of isotope exchange that is difficult to achieve, especially for materials with low diffusivities and sluggish reaction kinetics such as diamond and other carbons[5]. In case of nanoparticles, these difficulties become rather insuperable because proper isotope exchange experiments are barely possible due to grain growth effects such as Ostwald ripening, surface adsorption etc. In this light, development of non-perturbative methods estimating equilibrium stable isotope fractionation factors are critically important for nanoparticles.

Nuclear Resonant Inelastic X-ray Scattering (NRIXS) is one of few analytical methods capable to address relevant properties of nanoparticles. The partial or projection

(on vibrations of a chemical element of interest) Phonon Density of States (PDOS) obtained in NRIXS experiments allows calculation of the reduced isotopic partition function ratio (β-factor), the quantity controlling thermodynamic properties of appropriate isotopologues of a solid[6,7]. However, NIXRS is not readily applicable to light elements such as C, since it is observed only for elements having a Mössbauer isotope and its application is thus limited by heavier elements. This is also true for Mössbauer spectroscopy, another method allowing estimation of β-factors[8].

For single-element substances, the β-factors can be calculated from heat capacity data at elevated temperatures[9,10]. However, this approach cannot be applied to diamond nanoparticles due to lack of appropriate heat capacity data (for details see section "INS-derived β-factor and heat capacity for bulk diamond"). Fortunately, one can calculate the β-factors for single-element solids based on the complete DOS, since projection of the PDOS on vibrations of the chemical element of interest is not needed in this case. The calculation approach is mainly analogous to that previously applied to the Fe-metal PDOS from NRIXS[7,11].

In this contribution, we show that Inelastic Neutron Scattering (INS) is a promising experimental method for evaluation of isotopic properties. In particular, this approach is extremely useful for estimation of the β-factors of nanoparticles, where proper isotope exchange experiments are barely possible. Using experimental INS spectra, we evaluate the β-factors and heat capacity of nanodiamonds and estimate the size effect on these quantities.

2. **Samples and experimental details**

INS spectra were obtained for diamond powders with grains of markedly different sizes. Nanodiamonds were represented by milled High-Pressures – High Temperature synthetic diamonds with sizes (according to Dynamic Light Scattering) of 170 nm and 40 nm and by synthetic detonation nanodiamond (DND) with a grain size of 5 nm (data of X-ray diffraction, Small-Angle Scattering, and Transmission Electron Microscopy). As a reference sample, powder of synthetic diamond with grain sizes between 10 and 50 microns (denoted as macrodiamond) was used.

The powders with masses 2-4 g were placed in Al foil cylinders and measured at IN4C instrument at the Institute Laue-Langevin (ILL, Grenoble, France) at temperatures between 150 and 500 K in vacuum. The IN4C instrument is a time of flight spectrometer mounted on the thermal source at the High Flux Reactor of the ILL. A graphite monochromator selects neutrons with the desired wavelength from the white neutron beam. The incident beam is subsequently transformed into sharp neutron pulses using a Fermi chopper. Neutron energy after scattering from the sample is analysed by measuring the time it takes for a neutron to fly over a calibrated distance determined by the sample-to-detector distance. The instrument was set up as to use incident neutron wavelength of 2.4 Å in up-scattering mode (e.g. Anti-Stokes scattering). In this mode, the maximum frequency attainable depends on the temperature of the sample. At 150 K, one can easily derive the phonon spectrum from ~0.5 meV up to 50 meV providing the scattering of the sample is large enough. The unavoidable deterioration of the energy resolution with increasing energy transfer (resolution is 0.7 meV at elastic scattering, 1 meV at 10 meV, 1.4 meV at 20 meV and 2.9 meV at 40 meV - FWHM) can be minimized by time focusing in the inelastic range[12], a condition requiring that the Fermi chopper spins at high speed. We chose a Fermi speed of 17000 RPM and conditions so that frame overlap could be avoided. The measurements were corrected for the scattering of the sample holder and normalized to vanadium monitor. The signal was transformed into the so-called generalized density of states G(E) in the framework of the incoherent approximation13, after proper averaging of the scattered signal over the wide scattering angle (120°) provided by the IN4C multidetector. Details of an analysis of neutron scattering data can be found elsewhere[14]. Here we mention that G(E) is defined as:

$$G(E) = \sum_{\mu=1}^{N} c_\mu \frac{\sigma_\mu}{M_\mu} g_\mu(E) \qquad (1)$$

where the subscript µ runs over all atoms in the sample unit cell, $c_\mu$, $\sigma_\mu$, $M_\mu$, and $g_\mu(E)$ are the concentration, total scattering cross-section, mass, and partial phonon density of states (contribution of atom µ to the phonon density of states), respectively, for the µ-th atomic species. For monoatomic samples, G(E) is directly proportional to the phonon density of state g(E).

To reduce contribution of adsorbed water the samples were first vacuum-heated to 500 K and then gradually cooled. Though 500 K is insufficient for complete water desorption from 5 nm nanodiamonds[15,16], our INS spectra show some contributions of the adsorbed hydrogen. Importance of removing the adsorbed hydrogen and its influence on the β-factor and heat capacity are discussed below.

**Theory.**

**Calculating the β-factor and heat capacity from the PDOS.**

The reduced isotopic partition function ratio or β-factor is the main concept of the stable isotope thermodynamics controlling isotopic behavior of a substance (phase) in equilibrium processes. The equilibrium isotope fractionation factor between compounds A and B is related to the β-factors as:

$$\alpha_{A/B} = \beta_A / \beta_B \quad \text{or} \quad \Delta_{A-B} \approx \ln\alpha_{A/B} = \ln\beta_A - \ln\beta_B \qquad (2)$$

where $\beta_A$ and $\beta_B$ are the β-factors of compounds (phases) A and B, respectively, and $\Delta_{A-B}$ is the equilibrium isotopic shift between A and B phases.

In vast majority of cases, the β-factor is defined by differences in the vibration (phonon) spectra due to isotope substitution. In the harmonic approximation, the β-factor is expressed by the following equation[17,18]:

$$\ln\beta = \sum_{i=1}^{3N-6}\left(\ln\frac{\sinh(0.5u_i)}{\sinh(0.5u_i^*)} - \ln\frac{u_i}{u_i^*}\right) \qquad (3)$$

where $u \equiv h\nu/k_BT$ is the dimensionless frequency; $\nu$ is the normal (phonon) frequency; T is the absolute temperature; h and $k_B$ are the Planck and Boltzmann constants, respectively; superscript * defines, hereafter, quantities relating to a rare ($^{13}$C in case of carbon) isotopologue; the summation is over all 3N-6 vibration modes of a non-linear molecule. In case of solids, it is convenient to rewrite Eq. (3) as:

$$\ln\beta = 3N\int_0^{e_{max}}\left(\ln\frac{\sinh\left(0.5\dfrac{e}{k_BT}\right)}{\sinh\left(0.5\dfrac{e^*}{k_BT}\right)} - \ln\frac{e}{e^*}\right)g(e)\,de \qquad (4)$$

where $e = h\nu$ is the phonon energy; $g(e)$ is the PDOS of the main isotopologue ($^{12}$C in our case) and the integral is taken over the entire phonon spectrum from 0 to the maximal phonon energy ($e_{max}$). In Eq. (4), the PDOS is normalized to unity as following:

$$\int_0^{e_{max}} g(e)de = 1 \tag{5}$$

It is obvious that knowledge of the phonon energies (frequencies) for both isotopologues is required for calculation of the β-factor. Fortunately, in case of the single-element substances the normal harmonic frequencies of isotopologues are related by a simple equation:

$$\frac{e^*}{e} = \frac{v^*}{v} = \left(\frac{m}{m^*}\right)^{0.5} \tag{6}$$

in which $m$ and $m^*$ are masses of abundant and rare isotopes of interest (here masses of $^{12}$C and $^{13}$C), respectively. Substituting Eq. (6) into Eq. (4), one gets ($N = 1$ for single-element solids):

$$\ln \beta = 3 \int_0^{e_{max}} \left[ \ln \frac{\sinh\left(0.5 \frac{e}{k_B T}\right)}{\sinh\left(0.5 \left(\frac{m}{m^*}\right)^{0.5} \frac{e}{k_B T}\right)} - 0.5 \ln \frac{m^*}{m} \right] g(e)de \tag{7}$$

Equation (7) allows calculation of the β-factor of bulk and nanoparticle single element solids in the harmonic approximation.

The PDOS provides also a means for calculation of the heat capacity of bulk and nanoparticulate non-magnetic insulators. Heat capacity per mole can be calculated by averaging the Einstein equation for heat capacity of a single harmonic oscillator over the entire phonon frequency spectrum using the PDOS[19,20]

$$C_v = 3RN \int_0^{e_{max}} \left(\frac{e}{k_B T} \Big/ \exp\left(\frac{e}{k_B T}\right) - 1\right)^2 \exp\left(\frac{e}{k_B T}\right) g(e)de \tag{8}$$

where R is the universal gas constant; N = 1 in the case of single-element solids. Equation (8) is applicable both to nanostructured and bulk substances. We will use Eqs. (7) and (8) for evaluation of the size effect on the β-factor for diamond and its heat capacity, respectively.

# Extraction of the $^{12}$C component from the diamond PDOS

In our INS experiments diamond comprising natural mixture of $^{12}$C and $^{13}$C isotopes was used, whereas Eq. (7) for β-factor contains the PDOS of isotopically pure $^{12}$C diamond. We note that using first order of the thermodynamic perturbation theory it was shown that for all elements except hydrogen isotopic mixtures in solids can be considered ideal with accuracy sufficient (better than 1%) for analysis of isotopic effects at room and higher temperatures[21]. Assuming ideal isotope mixture one can represent the PDOS in diamond as a sum of components of pure $^{12}$C and $^{13}$C isotopes, weighted according to their abundances:

$$g_{nat}(e) = rg(e) + r^* g^*(e) \tag{9}$$

where $g_{nat}(e)$ is the PDOS of the natural diamond; r is the isotope abundance; superscript * denotes quantities related to heavy $^{13}$C isotope as before, i.e. r = 0.989; r* = 0.011.

One can express the $^{13}$C diamond PDOS through the $^{12}$C PDOS taking into account relation (6):

$$g^*(e) = g\left(e\sqrt{\frac{m^*}{m}}\right)\sqrt{\frac{m^*}{m}} \tag{10}$$

Substituting Eq. (10) into Eq. (9), one gets:

$$g(e) = \frac{1}{r} g_{nat}(e) - \frac{r^*}{r} g\left(e\sqrt{\frac{m^*}{m}}\right)\sqrt{\frac{m^*}{m}} \tag{11}$$

Since $r^*/r \ll 1$, the second term in the right-hand side of Eq. (11) is small in comparison to the first one. This provides a following iterative schema for calculation of $g(e)$:

$$g_0(e) = g_{nat}(e)$$
$$g_i(e) = g_{nat}(e) - r^*\left[g_{i-1}\left(e\sqrt{\frac{m^*}{m}}\right)\sqrt{\frac{m^*}{m}} - g_{i-1}(e)\right] \tag{12}$$

where subscripts near g indicate number of iteration. Tests of the iterative schema (12) have shown that already the first iteration provides sufficient accuracy in the PDOS. For this reason, one can rewrite Eq. (12) as:

$$g(e) = g_{nat}(e) - r^* \left[ g\left(e\sqrt{\frac{m^*}{m}}\right)\sqrt{\frac{m^*}{m}} - g_{nat}(e) \right] \quad (13)$$

Equation (13) is valid for phonon energies lower than the upper limit of the $^{13}$C diamond, i.e., $e \leq e_{max}(m/m^*)^{0.5}$. In the range between the upper phonon energy limits for pure $^{13}$C- and $^{12}$C-diamond, the PDOS for natural and $^{12}$C diamond are the same.

Consequently, one can write:

$$g(e) = g_{nat}(e) - r^* \left[ g\left(e\sqrt{\frac{m^*}{m}}\right)\sqrt{\frac{m^*}{m}} - g_{nat}(e) \right] \quad \text{at } 0 \leq e \leq e_{max}\sqrt{\frac{m}{m^*}}$$

$$g(e) = g_{nat}(e) \quad \text{at } e_{max}\sqrt{\frac{m}{m^*}} < e \leq e_{max} \quad (14)$$

Another method of extraction of the β-factor from isotope mixture PDOS based on assumption of ideality of the isotope mixture was established in ref. 22. They introduced a factor $\lambda_m$ to address quantum correction to classical kinetic energy for an isotope with mass m present in the mixture. As a consequence, one should multiply the lnβ, calculated from the PDOS of the isotopic mixture, by the factor $\lambda_m$ in order to obtain the corrected value of the β-factor. The following expression for the $\lambda_m$ takes place for a chemical element consisting of n isotopes[22]:

$$\lambda_{m_j} = \left[ \sum_{i=1}^{n} r_i \frac{m_j}{m_i} \right]^{-1} \quad (15)$$

where $m_i$ and $r_i$ are mass and abundance of the i-th isotope.

In the case of carbon, the $\lambda_{12}$ is:

$$\lambda_{12} = \left[ r + r^* \frac{m}{m^*} \right]^{-1} = 1.000849$$

One can see that for diamond the correction to the carbon lnβ is small.

## Results

## INS spectra of nanodiamonds.

PDOS for nanodiamonds recorded at 300 K are shown on fig. 1 in comparison with that for macrodiamond crystal. Although shapes of the curves are generally similar for different diamond samples, relative intensities of spectral features are clearly sample-dependent. Despite preliminary heating to 500 K, adsorbed hydrogen species contribute to the INS-derived PDOS of 5 nm and 40 nm nanodiamonds. They are manifested as broad humps below 60 meV leading to deviation of the PDOS curves from the parabolic law at low energies (Fig. 1). This component may significantly increase the heat capacity and reduce the β-factor of nanodiamond. They may be important since surface hydrogenation of nanoparticles is virtually inevitable even in outer space environment (e.g. ref 23). Below we estimate the effect quantitatively.

It is well known that primary particles of 5 nm nanodiamonds tend to form very stable aggregates due to Van-der-Waals interaction between the faces[24]. We have measured INS-derived spectra for two samples with aggregate sizes of 30 and 200 nm according to Dynamic Light Scattering. As expected, the INS spectra of these aggregates are identical within the Fig. 1. measurement error and contributions of the adsorbed H-species do not differ despite possible variations in porosity.

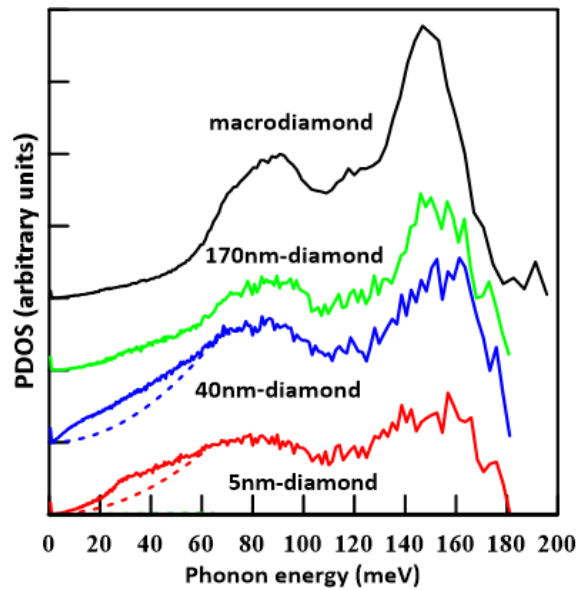

Fig. 1. Experimental PDOS obtained in INS experiments. Dashed lines show the parabolic dependence for the PDOS.

## INS-derived β-factor and heat capacity for bulk diamond

Prior to estimating differences in β-factors and heat capacities between macro- and nanodiamonds, it is reasonable to test the approach described in section "Theory" and to compare INS-derived β-factor and heat capacity values of macrodiamond (this study) with those from previous investigations.

As it follows from Fig. 2, our evaluation somewhat underestimates the β-factor for bulk diamond comparing to those from refs. 9 and 25, which agree to each other. The density functional theory (DFT) approach was used to calculate the PDOS of the bulk diamond[26]. The β-factor derived from the calculated PDOS by is also in a good agreement with estimations from refs. 9 and 25 (Fig. 2). The most recent direct β-factor calculation[27] based on DFT approach agrees with our result.

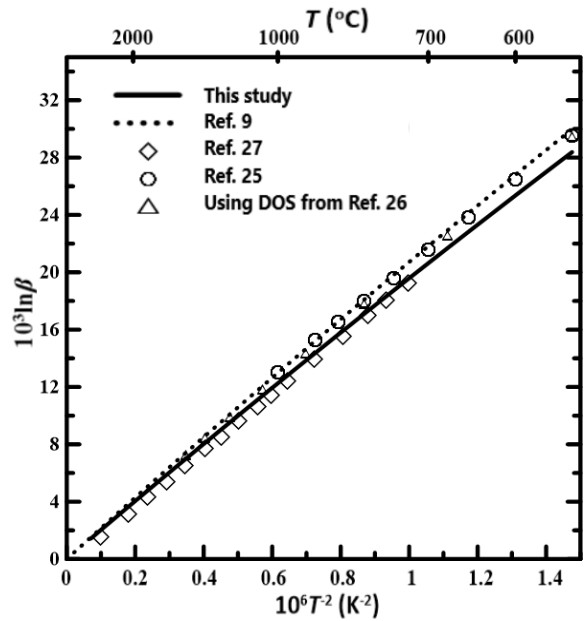

Fig. 2. Comparison of present and previous evaluations of the β-factor for bulk diamond.

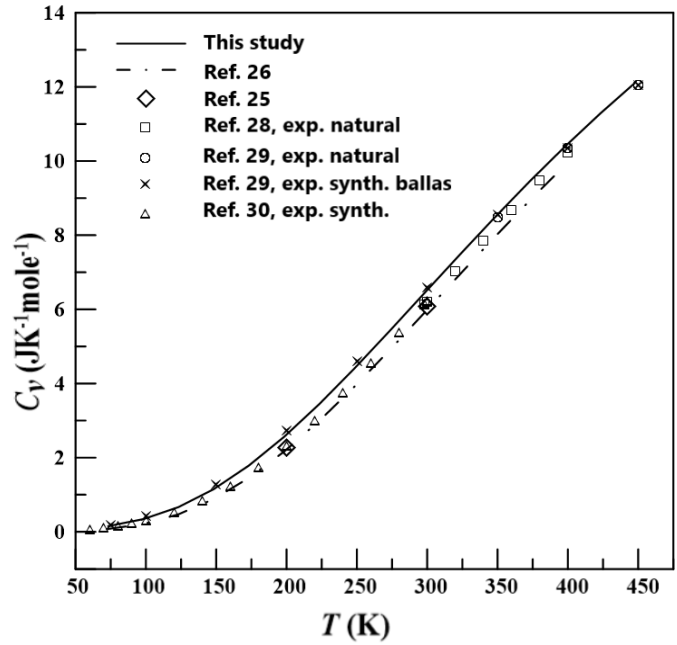

Fig 3. Comparison of present-study INS-derived heat capacity with theoretical and experimental data of previous studies.

Both theoretical and experimental data are available for heat capacity of diamond (Fig. 3). The theoretical predictions from ref. 25 based on dynamic lattice approach are in good agreement with the DFT-based calculation from ref. 26. In turn, these data agree well with measurements of heat capacity of natural and synthetic diamonds[28-30]. Our data give somewhat higher values than these theoretical and experimental results, but agree well with experimental measurements for synthetic ballas-type diamond containing metal inclusions[29] (Fig. 3). In order to avoid problems with quality of diamond samples and reveal size effects, we consider difference between macro- and nanodiamond in subsequent sections.

**Heat capacity of nanodiamond**

Results of heat capacity calculations of nanodiamonds from the INS-derived PDOS (Fig. 1) using harmonic Eq. (8) are shown in Fig. 4. The heat capacities of the nanodiamonds exceeds those of the bulk diamond at all temperatures. The excess is the highest between ~200 – 300 K and tends to zero at low and high temperature limits, where harmonic heat capacity at constant volume approaches zero and 3R (Dulong-Petit law), respectively (Fig. 5). One can see that the INS-derived heat capacity of 170 nm nanodiamond does not deviate significantly from that of bulk diamond. The difference is close to that between the theoretical prediction from ref. 26 and the present INS-based evaluation. The heat capacities of 5- and 40-nm nanodiamonds do not differ much from each other, but deviate significantly from that of bulk diamond (Fig. 5). These deviations and experimental calorimetry data[31] match in sign, but differ somewhat in magnitude. Some discrepancy between our and data from ref. 31 can be attributed to different amount and composition of impurities adsorbed on surfaces of nanodiamonds used in both experiments.

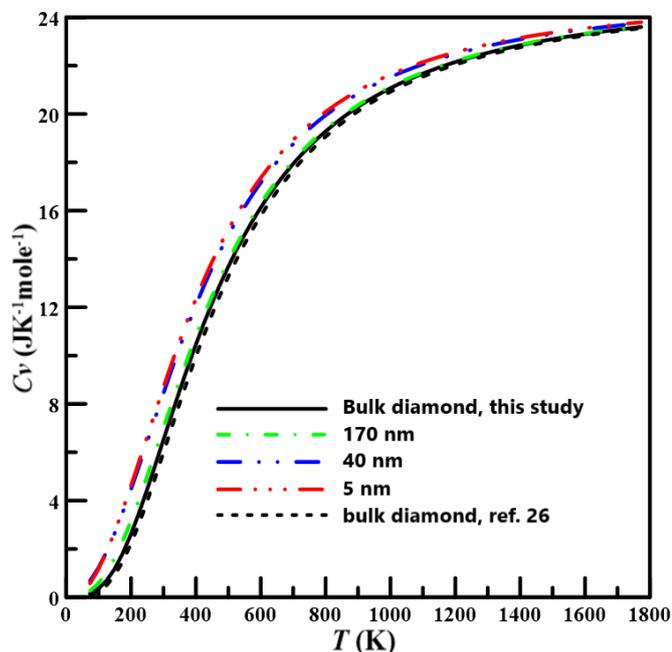

Fig. 4. INS-derived heat capacities at constant volume. Theoretical prediction from ref. 25 is shown for comparison (see text).

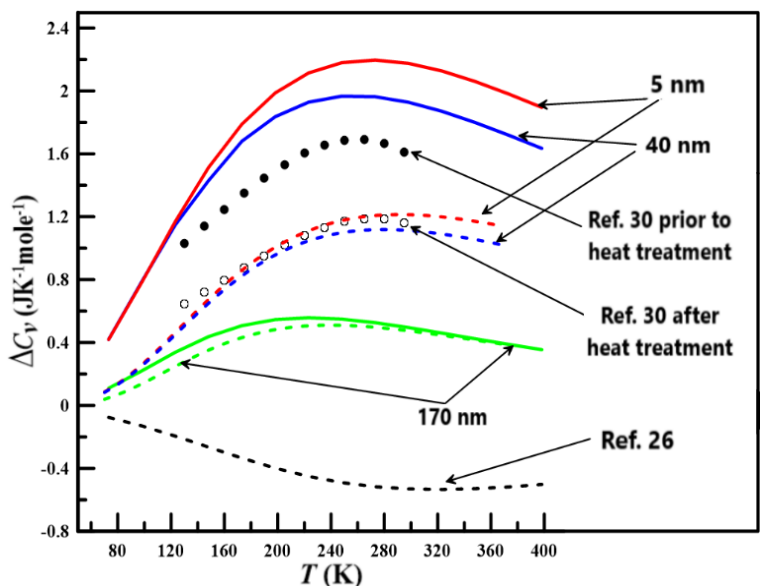

Fig. 5. The difference in heat capacity between nanodiamonds and bulk diamond calculated from INS-derived PDOS (solid curves) and modified PDOS (dashed lines). Calorimetry measurements from ref. 31 and the theoretical prediction from ref. 26 PDOS are presented for comparison.

As noted in the section "INS spectra of nanodiamond", H-containing impurities adsorbed on a nanoparticle surface contribute into the PDOS between ~40 – 70 meV and are responsible for deviation of the nanodiamond PDOS from the parabolic law (Fig. 1). The PDOS of bulk diamond follows the parabolic law up to ~66 meV. Assuming that both nano- and bulk diamond obey the parabolic law in the same range (Fig. 1), one can exclude contribution of the surface impurities into nanoparticles PDOS. Thus, heat capacity calculations for nanodiamonds based on PDOS with recovered the parabolic law at the low phonon energy range (Fig. 1) may provide estimating deviations of nanodiamond heat capacities relative to that of bulk diamond caused by phonon size effect (phonon confinement) on the PDOS, rather than the surface impurity influence. We compare these results with experimental data from ref. 31 who studied the effect of surface impurities on heat capacity of nanodiamond using calorimetry both prior to and after heating to 1000 K in vacuum (Fig. 5). The difference in Cv between the heat-treated nanodiamonds and bulk diamond are shown in Fig. 5 along with our estimates of the phonon confinement effect. A good agreement between the calorimetric and INS-derived phonon confinement effect in heat capacity of nanodiamonds is clear. Exact explanation of differences at low (<150 K) temperatures is yet lacking, but most likely, it is caused by impurities remaining after the heat treatment and/or high uncertainties in calorimetric measurements at small values of heat capacity in ref. 31. This agreement supports the suggestion that the non-parabolic behavior of the PDOS at low energies results from surface impurities. From Figs. 4 and 5 it follows that for 5- and 40 nm nanodiamonds about 50% of difference of heat capacity with bulk diamond stems from surface impurities. For the 170 nm diamond the influence of the surface shell is less important.

**β-factor for nanodiamond**

PDOS obtained in INS experiments (Fig. 1) was used for calculation of the β-factors of nanodiamonds using mathematical formalism developed in section "Theory"; the results for bulk diamonds are presented in section "INS-derived β-factor and heat capacity for bulk diamond". Results of the calculations are shown in Fig. 6 as temperature dependence of the equilibrium isotopic shifts (Δ) between bulk diamond and nanodiamonds: $\Delta$ (‰) ≈ $10^3 \ln\beta_{(bulk\ diamond)} - 10^3 \ln\beta_{nanodiamond}$. The theoretical formalism is valid for single-element solids. This condition is fulfilled for bulk diamond, where influence of the surface impurities on the β-factor is obviously negligible. The situation is less obvious in case of nanoparticles, when fraction of non-carbon surface atoms may be significant. For this reason, direct calculation of isotopic shifts (Fig. 6) may be compromised. As it follows from the heat capacity consideration in the preceding section, use of the modified PDOS (with recovered parabolic law at the low-energy range, Fig.1), reduces the surface effects on PDOS of nanoparticles. One can expect that calculations using the modified PDOS allow estimation of isotopic shifts between nano- and bulk diamond caused by differences in phonon behavior. Such calculations are shown in Fig. 6 as dashed lines. As for the case of heat capacity, the phonon confinement effects on the β-factor for nanoparticles is significant. The isotopic shifts between 5 nm and 40 nm-nanodiamonds and bulk diamond exceed that between diamond and graphite.

Somewhat surprisingly, the 40 nm nanodiamond sample deviates from the larger grains. We do not have unambiguous explanation of this phenomenon since 40 nm is still a relatively large grain with negligible fraction of surface-bound atoms and similarities with macroscopic grains are expected. The observed deviations are most likely explained by the process of preparation of the 40 nm nanodiamonds by mechanical grinding of synthetic macrodiamonds. It is known that nanodiamonds with sizes exceeding 10-20 nm prepared by this approach are often highly anisometric and flattened: whereas their lateral sizes reach several tens of nanometers (and are responsible for the size observed by Dynamic Light Scattering), the thickness can be considerably smaller, 10-20 nm[32].

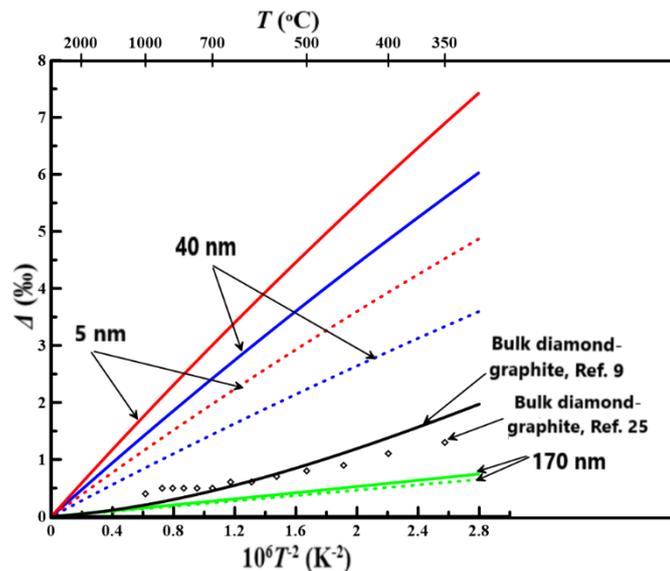

Fig. 6. Isotopic shift between bulk diamond and nanodiamonds. Solid lines relate to calculations using the INS-derived PDOS. Dashed lines correspond to the PDOS corrected for deviations from the parabolic law due to surface impurities (see text for details). The isotopic shift between diamond and graphite is shown for comparison.

**Applicability of different methods for calculation of the β-factor**

In the case of a single element solid, the PDOS and the projection PDOS coincide with each other. Thus, measurement of the PDOS for diamond provides perfect opportunity to test different techniques using for derivation of the β-factor from the projection PDOS and/or its moments obtained in NRIXS experiments[6,7,33-36]. The equation expressing the β-factor in terms of the kinetic energy of the nucleus of interest forms the basis for these calculation approaches[8,37].

$$\ln \beta = \left(\frac{K}{RT} - \frac{3}{2}\right)\frac{\Delta m}{m^*} \qquad (16)$$

where K is the kinetic energy (per one gram-atom) of the nucleus upon isotopic substitution in the main isotopologue; $\Delta m = m^* - m$. For diamond, $\Delta m$ is the difference between masses of $^{13}C$ and $^{12}C$ isotopes. Equation (16) is valid in the first order of the thermodynamic perturbation theory (TPT) and does not require knowledge of any thermodynamic quantities of the rare isotopologue. The PDOS (the projection PDOS in a general case) is required for calculating the kinetic energy of a energy and the virial theorem, in harmonic approximation one gets for the kinetic energy of the nucleus of interest[6,7]: nucleus upon isotopic substitution. Using the Einstein equation for harmonic oscillator

$$K = \frac{3}{2}RT \int_0^{e_{max}} \left[\frac{e}{k_BT} \bigg/ \left(\exp\left(\frac{e}{k_BT}\right) - 1\right) + \frac{0.5e}{k_BT}\right] g(e)de \qquad (17)$$

and taking into account Eq. (16) gets for the β-factor:

$$\ln \beta = \frac{3}{2}\left\{\int_0^{e_{max}} \left[\frac{e}{k_BT} \bigg/ \left(\exp\left(\frac{e}{k_BT}\right) - 1\right) + \frac{0.5e}{k_BT}\right] g(e)de - 1\right\}\frac{\Delta m}{m^*} \qquad (18)$$

Another approximate method for calculating the β-factor expresses the β-factor via moments of the PDOS[33,34] and named as the general moment (GM) method[34]. The equation of the GM method can be derived applying the Thirring expansion to equation for kinetic energy[19,33,38, ‡]:

$$\ln \beta = \frac{\Delta m}{m^*}\left(\frac{K}{RT} - \frac{3}{2}\right) = \frac{3}{2}\frac{\Delta m}{m^*}\sum_{i=1}^{\infty}(-1)^{i-1}\frac{B_{2i}}{(2i)!}\frac{E_{2i}}{(k_BT)^{2i}} = \frac{3}{2}\frac{\Delta m}{m^*}\sum_{i=1}^{\infty}(-1)^{i-1}\frac{B_{2i}}{(2i)!}\left(\frac{\theta_{2i}}{T}\right)^{2i} \qquad (19)$$

where $B_{2i}$ is the 2i-th Bernoulli number, $E_{2i}$ is the defined as $E_{2i} \equiv \int_0^{e_{max}} e^{2i} g(e) de$ and $\theta_{2i} = \sqrt[2i]{E_{2i}}/k_B$ is the characteristic temperature expressing the 2i-th moment of the PDOS in Kelvins. This series converges at $T > e_{max}/2\pi k_B$ where $e_{max}$ is the blue limit of the PDOS. The GM method is important for evaluation of the β-factors from NRIXS, since the moments of the projection PDOS can be directly obtained in these experiments[34,35,40,41].

Comparison of the Bigeleisen and Mayer (B-M) equation with the approximate methods is shown in Fig. 7 using dimensionless abscissa: $(\theta_2/T)^2$. The first-order of the TPT provides a good approximation to the B-M equation in the whole temperature range important for geochemical applications (Fig. 7). It always underestimates the exact value of the β-factor, since the second-order of the TPT correction to the β-factor is positive[21]. Inasmuch as Eq. (19) represents expansion of the kinetic energy into the alternating series, consideration of the odd number of terms in expansion (Eq. 19) overestimates the lnβ given by the first-order of the TPT. Consideration of the even number of the terms underestimates it. Keeping odd number of terms in series (Eq. 19), one can get better approximation to the B-M equation than that provided by the first-order of the TPT at high temperatures (Fig. 7b). In this case, the deviations of approximate β-factor values from the exact ones are very small.

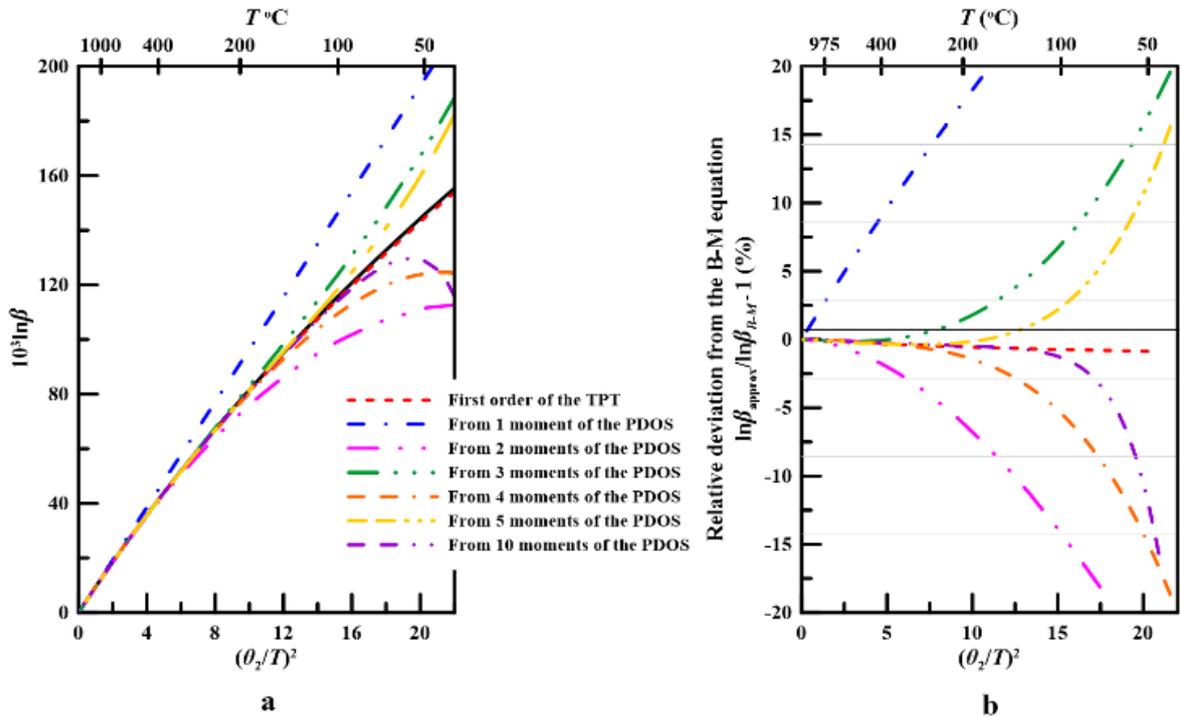

Fig. 7. Comparison of the approximate methods of the β-factor calculation with the Bigeleisen and Mayer (B-M) equation based on the diamond PDOS.

At lower temperatures, the use of the GM method results in large errors even despite large number of the terms in series (Eq. 19) are taken into account. Fig. 7b represents dependence of the relative deviation of approximate $\ln\beta$ values on dimensionless ratio $(\theta_2/T)^2$. Graphs in Fig. 7b depend on the form PDOS (relations between even moments of the PDOS) and are independent from magnitudes of the kinetic energy, moments, isotope masses, etc. When $(\theta_2/T)^2 \approx 20$ the magnitude of the relative deviations exceeds 10 % even though 10 terms of series in Eq. 19 are accounted for. For diamond $\theta_2 = 1454.7$ K and significant deviations are observed at temperatures exceeding room temperature. For the iron sublattices PDOS, for instance, $\theta_2$ does not exceed 600 K and the GM method provides a good approximation for geochemistry applications keeping 1, 2 or 3 terms of the series in Eq. (19)[7,34].

The Thirring expansion was first applied for calculation of heat capacity in refs. 42-44. The heat capacity data can be used for extraction of the β-factor from the moments of the PDOS[9,10,22]. However, the Thirring expansion provides good approximation to heat capacity of diamond at elevated temperatures, as it follows from calculations using the PDOS obtained in this study (Fig. 8). This does not allow estimating the β-factor for diamond nanoparticles using heat capacity data from ref. 31.

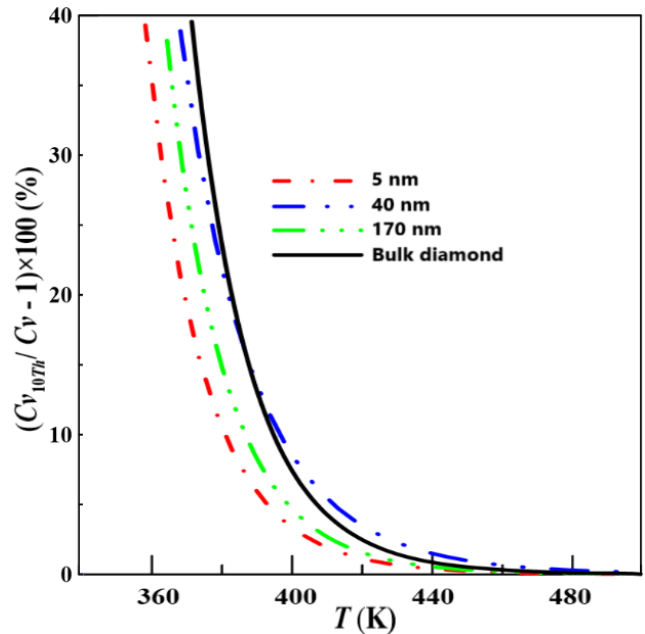

Fig. 8. Temperature dependence of the 10-order Thirring expansion error in the calculations of heat capacity for diamond nanoparticle. $Cv_{10th}$ is the heat capacity calculated using 10 even moments of the PDOS; $Cv$ is that calculated from the complete PDOS.

**Implications for natural nanodiamonds**

For nanostructured $Fe_{90}Zr_7B_3$ ribbons, the DOS of the grains does not change down to sizes of 2 nm and all deviations between the bulk material and the nanostructured one is ascribed to interfaces[45]. One might suggest that situation with nanodiamonds can also be explained in a similar scenario. In a popular model of structure of synthetic nanodiamond[46] it is postulated that a ND particle consists of a "perfect" diamond core enveloped into strained or onion-line $sp^2$-carbon (see ref. 47 for review). Surfaces of nanodiamonds from meteorites often contain a fraction of $sp^2$-bonded carbon[48], but their contribution is always rather minor; the same applies to synthetic nanodiamonds[47]. Whereas surface impurities and/or phase changes (here – partial $sp^3$-$sp^2$ conversion) do influence INS spectra, thermodynamic and isotopic properties of nanodiamonds, we show that contribution of these factors can be accounted for, and size-related isotopic effect does exist and may be important for isotopic patterns of nanoparticles.

Large-scale industrial process of nanodiamond (ND) synthesis involves detonation of oxygen-deficient substances in closed volumes[24]. $^{14}C$-labeled compounds were widely employed in studies of detonation ND synthesis[49], but these works obviously mostly provide chemical information about role of various functional groups in molecules of the explosives in formation of the diamond phase (see ref. 50 for review). In any case, the detonation synthesis is clearly far from equilibrium and thus is beyond the applicability of formalism of the current paper.

Meteoritic nanodiamonds are, perhaps, the most studied natural nanoparticles, at least among carbonaceous grains. Mechanism(s) of their formation remain debatable, but some variety of a Chemical Vapour Deposition process is the most likely one[51]. Whereas nitrogen-rich ND grains should have been formed very rapidly[52], growth of other particles could have been close to equilibrium and the isotopic size-effects described in this paper should be taken into account. Interestingly, nanodiamond fractions with different average sizes show small, but still significant variations[53] covering $\delta^{13}C$ values range between -26 and -32.8 ‰, i.e. the magnitude of the effect is comparable to the isotopic difference between macro- and nanodiamonds observed in the current work. Whereas differences between populations of nanodiamonds may reflect variations in growth conditions and/or origin, the present work shows that, at least in theory, equilibrium crystallization of grains with wide size distribution from a single source may produce measurable isotopic scatter.

**Conclusions**

Despite importance of nanoparticles for many natural and technological processes, getting reliable experimental thermodynamical data is a non-trivial task. In this paper, we propose a novel experimental method, which, together with developed mathematical formalism, allows determination of thermodynamical and equilibrium isotopic properties for nanoparticulate systems. We show that proper evaluation of contribution of surface contaminants or phases is possible.

For the first time considerable difference in equilibrium isotopic properties between nano- and macroparticles is demonstrated on example of diamond. It is shown that in equilibrium nanodiamonds are enriched in light carbon isotope in comparison with macroscopic diamond; thermal capacity of nanodiamonds exceed that of bulk diamond. In particular, the approach developed in the current work opens possibility to study thermodynamic properties of disordered and nanocarbons relevant for astrophysics and technology.

**Acknowledgements**

The study is partly supported by Russian Foundation of Basic Research (grants # 19-05-00865a and 13-05-91320-SIG-a).

**Notes and references**

‡ Authors of ref. 34 preferred another way of the derivation applying the first order of the perturbation theory to the Bigeleisen equation[39] expressing the β-factor via the difference in moments of the PDOS of isotopologues of interest. In addition, in ref. 34 the first term in Eq. (19) was expressed through the mean force constant following ref. 39.